\documentclass[aip,graphicx, reprint]{revtex4-2}
\usepackage{graphicx}

\usepackage{xcolor}
\usepackage{comment}
\usepackage{hyperref}
\usepackage{placeins}
\usepackage{afterpage}
\usepackage[normalem]{ulem}

\usepackage[T1]{fontenc}

\newcommand{\vone}[1]{{}}
\newcommand{\vtwo}[1]{{#1}}

\begin{document}


\title{Fast characterization of multiplexed single-electron pumps with machine learning} 



\author{N. Schoinas}
\thanks{These two authors contributed equally}
\author{Y. Rath}
\thanks{These two authors contributed equally}
\author{S. Norimoto}
\author{W. Xie}
\author{P. See}
\affiliation{National Physical Laboratory, Teddington, TW11 0LW, United Kingdom.}
\author{J. P. Griffiths}
\affiliation{Cavendish Laboratory, University of Cambridge, J. J. Thomson Avenue, Cambridge CB3 0HE, United Kingdom}
\author{C. Chen}
\affiliation{Cavendish Laboratory, University of Cambridge, J. J. Thomson Avenue, Cambridge CB3 0HE, United Kingdom}
\author{D. A. Ritchie}
\affiliation{Cavendish Laboratory, University of Cambridge, J. J. Thomson Avenue, Cambridge CB3 0HE, United Kingdom}
\author{M. Kataoka}
\affiliation{National Physical Laboratory, Teddington, TW11 0LW, United Kingdom.}
\author{A. Rossi}
\thanks{Authors to whom correspondence should be addressed:~alessandro.rossi@npl.co.uk; ivan.rungger@npl.co.uk}
\affiliation{National Physical Laboratory, Teddington, TW11 0LW, United Kingdom.}
\affiliation{
  Department of Physics, SUPA, University of Strathclyde, Glasgow G4 0NG, United Kingdom}

\author{I. Rungger}
\thanks{Authors to whom correspondence should be addressed:~alessandro.rossi@npl.co.uk; ivan.rungger@npl.co.uk}
\affiliation{National Physical Laboratory, Teddington, TW11 0LW, United Kingdom.}
\affiliation{Department of Computer Science, Royal Holloway, University of London, Egham, TW20 0EX, United Kingdom.}


\date{\today}

\begin{abstract}
  We present an efficient machine learning based automated framework for the fast tuning of single-electron pump devices into current quantization regimes.
  It uses a sparse measurement approach based on an iterative active learning algorithm to take targeted measurements in the gate voltage parameter space. When compared to conventional parameter scans, our automated framework allows us to decrease the number of measurement points by about an order of magnitude.
  This corresponds to an eight-fold decrease in the time required to determine quantization errors, which are estimated via an exponential extrapolation of the first current plateau embedded into the algorithm.
  We show the robustness of the framework by characterizing 28 individual devices arranged in a GaAs/AlGaAs multiplexer array, which we use to identify a subset of devices suitable for parallel operation at communal gate voltages. The method opens up the possibility to efficiently scale the characterization of such multiplexed devices to a large number of pumps.
\end{abstract}


\maketitle 


Single-electron pumps are nanoscale devices that can produce quantized macroscopic electric currents by clocking the transfer of individual electrons to an external periodic drive~\cite{Pekola_2013,Kaestner_2015,Kaneko_2016,Giblin2012TowardsAQ, Giblin2019EvidenceFU,Keller1996AccuracyOE, Janssen2000AccuracyOQ, Stein2017RobustnessOS, Zhao2017ThermalerrorRI}.
This device technology has been primarily developed to realize the practical implementation of the SI unit of current, the ampere, which since 2019 is defined by the fixed value of the elementary charge, $e$~\cite{brochure2019mise,janssen2014redefinition,Scherer_2019}.
The overarching goal is the experimental realization of devices generating a quantized current according to the relationship $I=n \, e \, f$, where $f$ is the periodic drive frequency and $n$ is an integer multiple of electrons transferred in a cycle.

The device operation requires a large degree of manual intervention to find the appropriate operation conditions in a large space of control parameters.
With the increasing need of operating multiplexed devices in a parallel configuration to generate usefully large quantized currents~\cite{maisi2009parallel, mirovsky2010synchronized, ghee2019parallelized}, the manual tuning of control parameters for each device becomes a significant bottleneck.
Since each pump has slightly different operating parameters, this severely limits the pace at which candidate devices can be screened.

To tackle this limitation, here we present a machine learning (ML) based framework that supports the automatic tuning of multiple single-electron pumps.
The use of ML for experimental control of quantum devices is becoming increasingly popular, as it may unleash significant speedups \cite{PRXQuantum.2.040324, schuff2024fully}, \vtwo{ for example by using pre-trained predictive models to reduce the number of required measurement points informed by a Bayesian approach~\cite{lennon2019efficiently}.}

Our framework automatically finds and characterizes the $n=1$ plateau in single-electron pumps as a function of control DC voltages \vtwo{without requiring pre-training with existing pump map patterns}.
We focus on the $n=1$ plateau since this is the region where the pumps are typically operated to achieve the best current quantization~\cite{Giblin2012TowardsAQ, Giblin_2017}.
We present an active learning (AL) sparse measurement (ALSM) framework in which measurements are obtained iteratively in a data-driven approach, which is designed to gain the necessary information from as few measurements as possible.
To this aim, the method needs to find the boundaries of the $n=1$ plateau and acquire sufficient data to perform an exponential extrapolation from the boundary to the center of the plateau, a technique used to estimate the quantization error beyond the measurement noise floor~\cite{extrapol}.

\begin{figure*}
  \centering
  \includegraphics[width=\textwidth]{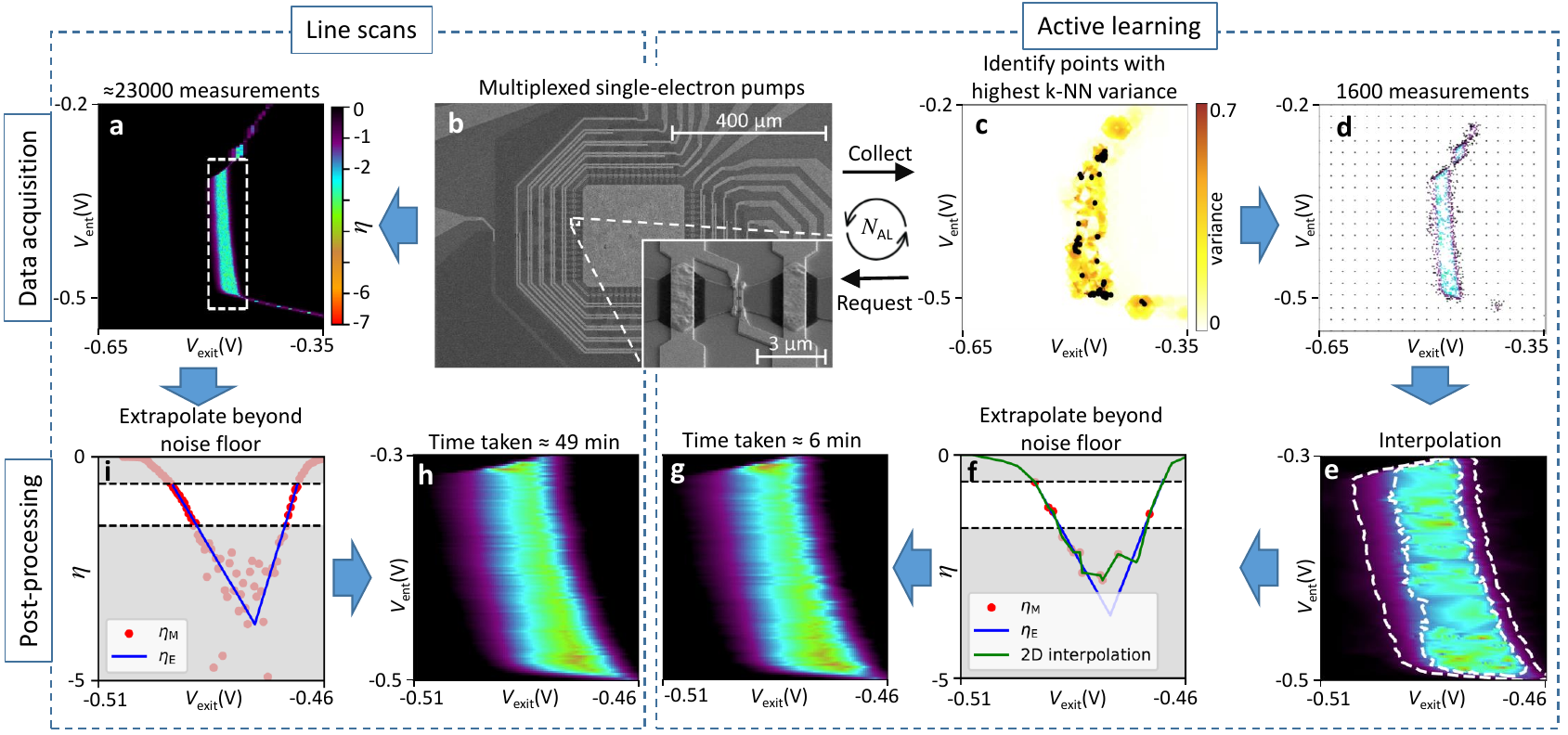}
  \caption{
    Schematic overview of the ALSM approach to characterize the $n=1$ plateau. a) $\eta_\mathrm{M}$, acquired through a PLS, as a function of $V_\mathrm{ent}$ and $V_\mathrm{exit}$ voltages, where a coarse scan over a wider parameter range is followed by a fine line scan around the $n=1$ plateau ($V_{\mathrm{exit}}$ coarse stepsize is $0.005 \, \mathrm{V}$, fine stepsize is $0.0005 \, \mathrm{V}$ ; $V_{\mathrm{ent}}$ stepsize is $0.002 \, \mathrm{V}$) b) scanning electron microscope (SEM) image of the 64-pump GaAs/AlGaAs multiplexer array; c) variance of $\eta_\mathrm{M}$ obtained with the k-NN active learning (AL) method at the $3^\mathrm{rd}$ iteration, where the black dots indicate the $N_\mathrm{meas}=60$ points that the k-NN method selects for measuring in the next iteration; d) sparse measured $\eta_\mathrm{M}$ points after $N_\mathrm{AL}=20$ AL iterations; e) interpolation of $\eta_\mathrm{M}$ from measurements in d) to a regular dense grid; the inner dashed curve encloses points with $\eta$ below the measurement noise floor $\eta_\mathrm{noise}$, while the outer dashed curve delimits the area of the $n=1$ plateau ($\eta_\mathrm{max}=-0.6$); f) ALSM and i) PLS $\eta$ as function of $V_\mathrm{exit}$ for $V_{\mathrm{ent}}= -0.464 \, \mathrm{V}$, illustrating the exponential fit (blue lines) of the $\eta_\mathrm{M}$ in a) (PLS) and e) (ALSM) for both sides of the plateau; the lower (upper) dashed line corresponds to $\eta_\mathrm{noise}$ ($\eta_\mathrm{max}$); for ALSM the fit is performed to the 2D interpolation data (green curve) rather than the measurement points (red dots); g) ALSM and h) PLS final extrapolated quantization error, $\eta_{E}(V_\mathrm{exit},V_\mathrm{ent})$, resulting from the the exponential fits (blue lines in f and i) for the whole 2D area. The color scale presented in a) is used for all 2D maps of $\eta$.
  }
  \label{fig:schematic}
\end{figure*}

We apply our protocol to a multiplexed array of GaAs/AlGaAs quantum dot pump devices~\cite{ShotaMultiplexer}, driven at a frequency of $f= 0.2 \, \mathrm{GHz}$, with a fixed external magnetic field of $B = 12.5 \, \mathrm{T}$ (Fig.~\ref{fig:schematic}b). The devices periodically trap and transfer a fixed number of electrons through a quantum dot (QD) with a single drive signal~\cite{kaestner2008}.
Such periodic drive is superimposed to a static DC voltage at the entrance barrier of the QD ($V_{\mathrm{ent}}$) to precisely clock the operation of the pump whilst the exit barrier is kept at a fixed DC voltage ($V_{\mathrm{exit}}$).
Depending on the setting for $V_{\mathrm{ent}}$ and $V_{\mathrm{exit}}$, the operation can be in the regime of quantized transfers taking place with specific number of electrons, or in the regime where the charge transfer is not quantized due to insufficient loading or incomplete emission~\cite{Kaestner_2015}.
To quantify the deviation from the $n=1$ quantization of the measured average electron number per cycle, $\langle n \rangle=I/ef$, we use the single-electron quantization error, defined as $\eta = \log_{10}(|\langle n \rangle -1|)$, so that a lower value of $\eta$ represents better current quantization.

The operation of our ALSM framework in comparison with a conventional line-scan method is presented in Fig.~\ref{fig:schematic}.
We establish a baseline with the traditional approach in which we scan $V_{\mathrm{exit}}$ for different $V_{\mathrm{ent}}$ values~\cite{howe2021single}, which  allows us to extract a heat map of $\eta$ over the parameter space. The grid scan consists of an initial coarse voltage scan over the full range of interest, followed by a higher resolution scan in the region around the $n=1$ plateau. We denote this protocol as plateau line scan (PLS). In Fig.~\ref{fig:schematic}a, we show the measured PLS $\eta$, which we denote as $\eta_\mathrm{M}$, as function of $V_\mathrm{exit}$ and $V_\mathrm{ent}$ for one pump. One can identify the primary large $n=1$ plateau at the center of Fig. \ref{fig:schematic}a (dashed rectangle). Additionally, the secondary $n=1$ plateaus obtained through incomplete emissions are visible above it. In the rest of the manuscript, we focus on the automatic characterization of the primary plateau.

A large fraction of the measurements in a PLS, for example the extended black regions, give redundant information. Our ALSM framework is designed to minimize the number of measurements by iteratively identifying the points that are expected to provide the largest information gain by using a k-nearest neighbor (k-NN) regressor~\cite{timbers2022data}.
This extracts a relationship between gate voltages and $\eta$ at a low computational cost~\cite{aly2015cost}, and is hence ideally suited to be included in the data acquisition loop. Its prediction for $\eta$ at unmeasured points is given by an average over already measured neighboring points, where neighbors are weighted according to their inverse distance.
At each iteration, we choose new points to be measured according to the variance in $\eta$ over their k-NN neighbors, giving a proxy metric that determines the uncertainty of the estimated $\eta$ values.
The first iteration consists of a very coarse scan with $N_\mathrm{coarse}$ measurements points over a large voltage area, followed by small batches of $N_\mathrm{meas}$ measurement points selected in each AL iteration, balancing additional time overhead to perform the k-NN prediction and data transfer versus keeping $N_\mathrm{meas}$ small, for a number of iterations, $N_\mathrm{AL}$.
At each iteration, we identify the $N_\mathrm{candidates}$ points that have the largest k-NN variance on a fine evaluation grid as potential candidates for a measurement in the next round, and sample the next measurement batch from these points with uniform sampling probability.
For the results presented here, we use $N_\mathrm{meas}=60$, $N_\mathrm{candidates}=100$, $N_\mathrm{coarse}=400$, $N_\mathrm{AL}=20$ and $k=4$~\cite{scikit-learn}.
Illustratively, we show the variance in the $3^\mathrm{rd}$ iteration of the AL cycle as a heat map in Fig.~\ref{fig:schematic}c, which also indicates the points selected for measurement in the following iteration as black dots. A large variance is typically found when $\eta$ varies strongly, or when measurement data is mostly absent in some region.
The 2D AL pump map for $\eta_\mathrm{M}$ after $N_\mathrm{AL}$ iterations is shown in Fig.~\ref{fig:schematic}d. The sparse measurement points are predominantly placed at the boundary regions of the $n=1$ plateau. When compared to the traditional PLS, which uses about $23,000$ measurements ($10,736$ for the coarse scan and $12,495$ for the fine scan), the AL framework only performs $1,600$ measurements $(N_\mathrm{AL} \times N_\mathrm{meas}+N_\mathrm{coarse}$), a reduction of more than an order of magnitude.

After completion of the AL cycle, we perform a two-dimensional piecewise linear interpolation between the measurement data~\cite{fukuda2004frequently, barber2013qhull, virtanen2020scipy}, allowing us to produce data on the same fine grid around the $n=1$ region used in the PLS. The results are shown in Fig.~\ref{fig:schematic}e. Within the $n=1$ plateau, it results in a smooth function until the data becomes noisy in the central area of the plateau with the highest current quantization accuracy. In these regions, the values of $\eta_\mathrm{M}$ are markedly affected by the measurement random uncertainty, given the limited averaging used ($20$ milliseconds). We assume this corresponds to the measurement noise floor for $\eta_\mathrm{M}$, and we denote it as $\eta_\mathrm{noise}$. Note that our AL method avoids taking a large number of measurements in this region, which would be dominated by noise.

To obtain estimates beyond $\eta_\mathrm{noise}$, we fit the $\eta$ with $\eta>\eta_\mathrm{noise}$ to an exponential approach of the current to the plateau from both sides using a regression analysis (Fig.~\ref{fig:schematic}i and f), which corresponds to a linear fit in $\eta$. For the PLS, we fit $\eta_\mathrm{M}$ directly (Fig.~\ref{fig:schematic}a), while for ALSM we fit the interpolated values (Fig.~\ref{fig:schematic}e).
Such an exponential function is a common phenomenological approximation for the approach to the plateau regions, whose actual physical behavior is masked by the limited measurement accuracy~\cite{extrapol}.
We denote the extrapolated quantization errors as $\eta_\mathrm{E}(V_\mathrm{exit},V_\mathrm{ent})$.
We determine the value of $\eta_\mathrm{noise}$ via ML density analysis~\cite{chen2017tutorial,wkeglarczyk2018kernel,scikit-learn} of the occurrence frequency, $p(\eta)$, of $\eta$ values across the full area of interest; $p(\eta)$ exhibits a peak where the data is dominated by noise, so that we set $\eta_\mathrm{noise}$ at the position of the local minimum of $p(\eta)$ above this peak (see Supplementary Materials for details).
We also set an upper threshold of $\eta_\mathrm{max}=-0.6$ for the $\eta$ included in the regression, since we consider points with larger values to be outside $n=1$ plateau region.

The obtained 2D function $\eta_\mathrm{E}(V_\mathrm{exit},V_\mathrm{ent})$ is the final output of our pipeline, which is shown in Figs. ~\ref{fig:schematic}h,g for PLS and ALSM, respectively. The ALSM heat map reproduces the reference PLS result well, and both have regions where the single-electron quantization error reaches values of $\eta_\mathrm{E} \approx -5$.
Whereas the PLS uses $\approx 23,000$ single point measurements resulting in a total time of $\approx 49$ minutes on average, the AL result is obtained with $1,600$ measurements, reducing the total time required to characterize the plateau to $\approx 6$ minutes.
This total time is made up of $\approx 5$ minutes taken for the experimental data acquisition (consisting of the bare measurement times, as well as additional control and communication overheads for interfacing with the experiment), an additional time of less than $4$ seconds in total for the identification of measurement points with the k-NN algorithm, and a total post-processing time of $\approx 38$ seconds.

\begin{figure}
  \centering
  \includegraphics[width=\columnwidth]{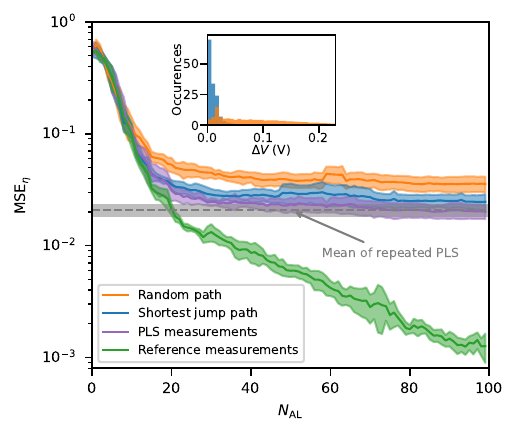}
  \caption{
  $\mathrm{MSE}_\eta$ of the extrapolated single electron quantization error $\eta_\mathrm{E}$ as function $N_\mathrm{AL}$, and where $\eta_\mathrm{E,ref}$ for the reference in Eq.~\ref{eq:mse} are data of a full dense scan over the final grid. The underlying ALSM data is obtained with a random (shortest jump) sequence measurement path for the orange (blue) curve. Solid lines and shaded areas denote the mean and standard deviation over five different realizations for each setting, respectively. The inset shows the normalized histogram of the step sizes taken in the random sequence and shortest jump paths. For comparison, the horizontal dashed line shows the averaged $\mathrm{MSE}_\eta$ for five analogous PLS. $\mathrm{MSE}_\eta$ where the AL measurement data at each AL round is replaced with the data of the reference (PLS) measurement for the same $(V_{\mathrm{exit},l},V_{\mathrm{ent},l})$ is shown in green (purple).
  }
  \label{fig:learning_curve}
\end{figure}

\begin{figure*}[htb!]
  \centering
  \includegraphics[width=\textwidth]{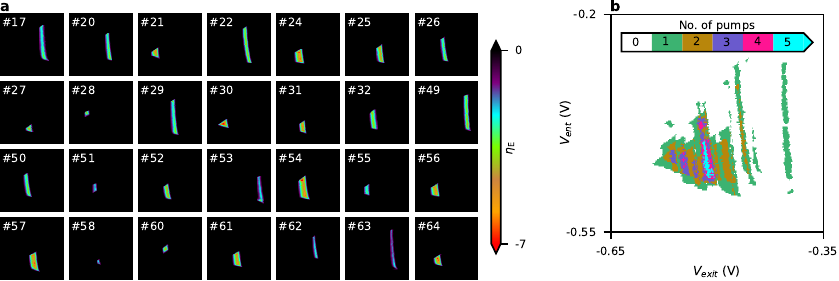}
  \caption{a) Active learning $\eta_\mathrm{E}(V_\mathrm{exit},V_\mathrm{ent})$ of 28 pumps of the multiplexed chip; voltage ranges are the same for all maps and shown in panel (b); b) number of overlapping pump regions with a quantization error $\eta_\mathrm{E}<10^{-3}$, showing that a maximum of five pumps could be operated in parallel with common gate voltages to within this quantization error.}
  \label{fig:overlap_with_heat maps_}
\end{figure*}

Having established the considerable speedup of our ALSM approach and the qualitative agreement with the PLS, we evaluate quantitatively the fidelity of the obtained $\eta_\mathrm{E}(V_\mathrm{exit},V_\mathrm{ent})$ when compared with line scan results. To quantify the fidelity between a given method and a reference measurement approach, we compute the mean squared error
\begin{equation}
  \mathrm{MSE}_\mathrm{\eta}=\frac{1}{N_\mathrm{p}}\sum_{l=1}^{N_\mathrm{p}}\left(\eta_{\mathrm{E},l}-\eta_{\mathrm{E,ref},l}\right)^2,
  \label{eq:mse}
\end{equation}
where $\eta_\mathrm{E,l} =\eta_\mathrm{E}(V_{\mathrm{exit},l},V_{\mathrm{ent},l})$, $N_\mathrm{p}$ is the total number of grid points in the final data, and the sum goes over all $(V_{\mathrm{exit},l},V_{\mathrm{ent},l})$ points in Figs.~\ref{fig:schematic}h,g where $\eta_{\mathrm{E,ref},l}$ is smaller than zero.
As reference data, we use $\eta_\mathrm{E}$ obtained for a separate \vtwo{single} full dense line scan with $N_\mathrm{p}$ measurements.

In Fig.~\ref{fig:learning_curve}, we present the results averaged over five \vtwo{separate} experiments of AL and PLS measurements, with an approximately one and a half hour time delay between experiments of the same type.
The $\mathrm{MSE}_\eta$ value for the repeated PLS data sets the lowest possible $\mathrm{MSE}_\eta$ (dashed line), and reflects the presence of measurement noise for the line-scan itself, as well as potential drifts of the device during measurements.
The orange curve is the $\mathrm{MSE}_\eta$ for a random sequence of measurement acquisition points within a given AL set of points. When increasing $N_\mathrm{AL}$, $\mathrm{MSE}_\eta$ initially decreases rapidly before it plateaus after about $N_\mathrm{AL}\approx 20$.

The fast early decrease of $\mathrm{MSE}_\eta$ is due to the progressively larger amount of information available for the interpolation between measurement points when a larger number of data points is included in the ALSM. When further increasing $N_\mathrm{AL}$ beyond $20$, the averaged $\mathrm{MSE}_\eta$ then converges to a value that is around a factor of two larger than the reference value set by repeated PLS.
For very large $N_\mathrm{AL}$, the main difference to a PLS approach lies not in the number of measured points but in the way these are acquired.
Whereas traditional line scans have a small constant step size between consecutive $(V_{\mathrm{exit},l},V_{\mathrm{ent},l})$ points, $\Delta V=\sqrt{\Delta V_\mathrm{exit}^2+\Delta V_\mathrm{ent}^2}$, potentially large and irregular jumps are taken in the ALSM approach. The histogram of the ALSM step sizes (inset of Fig.~\ref{fig:learning_curve}) shows that such jumps are randomly distributed over a range of about $0.2\,\mathrm{V}$.
We find that the averaged difference between the measured ALSM, $\langle n \rangle$, and the reference measurements increases as a function of $\Delta V$, which is also more pronounced around the edges of the plateau (see Supplemental Material for details).
This is likely due to the fact that large voltage steps can cause systematic errors for a given time-delay between measurements, since the measurement apparatus settling times for large $\Delta V$ increases.
Such memory effects can, for example, be due to the RC time constant of the low-pass filters inserted in the gate voltage lines protecting the gates from voltage surges, resulting in finite settling times when switching voltages.
Note that also the sequential stepping used in the line scans for the reference data may exhibit time correlations of the measurement data resulting in a spurious parameter space correlation of the data, which may be the cause for slight variations in positions of the plateau edges.
Part of the $\mathrm{MSE}_\eta$ may be attributed to a spurious correlation in the line scan data that is not present in the ALSM data due to the largely unbiased data acquisition path.

To mitigate effects of insufficient voltage settling between measurements, we therefore optimize the ALSM by ordering the measurement sequence to approximately minimize $\Delta V$ between measurements. As shown in the inset of Fig.~\ref{fig:learning_curve}, this optimization results in a large reduction in the range of step sizes (blue histogram). The resulting $\mathrm{MSE}_\eta$ (blue curve) is significantly reduced when compared to the random sequence results, and with increasing $N_\mathrm{AL}$ systematically converges to a value just slightly higher than the PLS baseline.

The question then arises how much of the $\mathrm{MSE}_\eta$ (if any) is solely introduced by the \vone{way the AL approach itself collects data}\vtwo{reduced set of measurement points acquired in the ALSM when compared to a PLS, separating it out from the noise and drift induced measurement fluctuations over time.
\vone{We can answer this by replacing the AL measurement data at each AL round with a sparse data set from the reference measurements for the same $(V_{\mathrm{exit},l},V_{\mathrm{ent},l})$ points queried in the AL experiment, and plot the results as green curve in Fig.~\ref{fig:learning_curve}.
This curve systematically decreases with increasing $N_\mathrm{AL}$, indicating that the AL approach itself can reach arbitrarily low target $\mathrm{MSE}_\eta$. When we replace the AL measurement data with its corresponding PLS data rather than the full line scan reference, the $\mathrm{MSE}_\eta$ systematically converges to the PLS baseline at around 60 AL rounds (purple curve).}
To this aim, we evaluate the $\mathrm{MSE}_\eta$ for a post-processed data set, where we keep the measurement points $(V_{\mathrm{exit},l},V_{\mathrm{ent},l})$ from our ALSM scan of the blue curve in Fig. \ref{fig:learning_curve}, but where we replace the measurement data itself with that of two different measurement sets:
\begin{itemize}
\item in the first, we replace the ALSM measurements for each of the five scans with those of the corresponding PLS at the ALSM $(V_{\mathrm{exit},l},V_{\mathrm{ent},l})$ points (purple curve in Fig.~\ref{fig:learning_curve}).
\item in the second, we replace the ALSM measurements for each of the scans with those of the reference line scan at the ALSM $(V_{\mathrm{exit},l},V_{\mathrm{ent},l})$ points. This gives MSE$_\eta$ for a hypothetical experiment where there are no differences in measurement values to the reference measurements for all $(V_{\mathrm{exit},l},V_{\mathrm{ent},l})$ (green curve in Fig.~\ref{fig:learning_curve}). 
\end{itemize}
The $\mathrm{MSE}_\eta$ for the first case (purple curve) systematically converges to the PLS baseline at around 60 AL rounds. On the other hand, for the second case (green curve) the $\mathrm{MSE}_\eta$ keeps decreasing with increasing $N_\mathrm{AL}$. This indicates that in absence of measurement fluctuations the AL approach itself can reach arbitrarily low target $\mathrm{MSE}_\eta$.
}

We showcase the robustness of our ALSM framework by using it to automatically characterize the $n=1$ plateau for pumps from a multiplexed chip with $64$ devices. Separate line scans show that a subset of $28$ pumps exhibits a plateau region, whereas the remaining ones do not exhibit well-defined quantization~\cite{ShotaMultiplexer}. The plateaus extracted with our ALSM approach are shown in Fig.~\ref{fig:overlap_with_heat maps_}a.
Our approach is able to extract a characteristic pump heat map for all the devices with reliable single-electron operation, highlighting the ability to faithfully find this ``needle in the haystack'' with only $1,600$ measurements per pump. Our algorithm estimates lowest quantization errors as low as $\eta_\mathrm{E} \approx-8.7$ across the $28$ pumps.
In Fig.~\ref{fig:overlap_with_heat maps_}b, we show how many pumps exhibit $\eta_\mathrm{E}<10^{-3}$ for a given $(V_\mathrm{exit},V_\mathrm{ent})$ parameter setting.
It allows to select both the parameter settings and the specific pumps that can be operated in parallel to reach this target $\eta$ \vtwo{below the identified noise floor} on each pump. For this multiplexer and measurement settings, our \vone{results}\vtwo{method} predict\vtwo{s} that a maximum of five pumps can be operated in parallel over extended parameter regions to obtain $\eta_\mathrm{E}<10^{-3}$ for each.
\vtwo{This is the result of a trade off. In fact, setting more stringent target quantization errors would lead to smaller suitable areas of operation for individual pumps, and this would, in turn, drastically reduce areas of overlap among pumps.}
Note that the use of communal gate voltages for parallel operation is needed for multiplexer device architectures for which some gate electrodes are shared among multiple pumps~\cite{ShotaMultiplexer}. For devices, where $(V_\mathrm{exit},V_\mathrm{ent})$ can be set independently for each pump, \vtwo{or for devices with communal gate voltages with an increased number of pumps, one can expect that a larger number of pumps could be operated in parallel for a given quantization accuracy}. \vtwo{O}\vone{o}ur approach allows for fast characterization and optimization of the $(V_\mathrm{exit},V_\mathrm{ent})$ parameters for \vone{each pump}\vtwo{a large number of pumps} to maximize the number of pumps that can be operated in parallel to within a target $\eta$.

In conclusion, we have developed, and demonstrated, an automatic machine learning based active learning sparse measurement protocol to characterize single-electron pumps for metrological applications. Our results enable time-efficient measurements by selecting meaningful data points to build quantization maps, as opposed to the conventional data acquisition approach relying on comprehensive exploration of the voltage parameter space by line scans. Our approach lends itself to be extended to other control parameter dimensions, such as drive signal amplitude and frequency, or magnetic field. Although our application focuses on the characterization of the first plateau region, the pipeline is directly transferable to higher level plateaus, making it a general prototype for the fully automated tuning of single-charge pumps. The demonstration that this AL-based framework can also be used for a large array of devices underlines its ability to streamline the experimental operation of single-electron pumping experiments towards parallelization.
\section*{Supplementary Material}
Details of the extrapolation beyond the noise floor, analysis of measurement errors as a function of voltage step size.
\begin{acknowledgments}
  The authors acknowledge useful discussions with S.P. Giblin.
  The authors acknowledge the support of the Innovate UK project AutoQT (grant number 1004359), and of the UK government Department for Science, Innovation and Technology through the UK National Quantum Technologies Programme.
  A.R. acknowledges support from a UKRI Future Leaders Fellowship (MR/T041110/1).
  The data that support the findings of this study are available from the corresponding authors upon reasonable request.
\end{acknowledgments}

\bibliography{main}

\begin{thebibliography}{31}%
\makeatletter
\providecommand \@ifxundefined [1]{%
 \@ifx{#1\undefined}
}%
\providecommand \@ifnum [1]{%
 \ifnum #1\expandafter \@firstoftwo
 \else \expandafter \@secondoftwo
 \fi
}%
\providecommand \@ifx [1]{%
 \ifx #1\expandafter \@firstoftwo
 \else \expandafter \@secondoftwo
 \fi
}%
\providecommand \natexlab [1]{#1}%
\providecommand \enquote  [1]{``#1''}%
\providecommand \bibnamefont  [1]{#1}%
\providecommand \bibfnamefont [1]{#1}%
\providecommand \citenamefont [1]{#1}%
\providecommand \href@noop [0]{\@secondoftwo}%
\providecommand \href [0]{\begingroup \@sanitize@url \@href}%
\providecommand \@href[1]{\@@startlink{#1}\@@href}%
\providecommand \@@href[1]{\endgroup#1\@@endlink}%
\providecommand \@sanitize@url [0]{\catcode `\\12\catcode `\$12\catcode `\&12\catcode `\#12\catcode `\^12\catcode `\_12\catcode `\%12\relax}%
\providecommand \@@startlink[1]{}%
\providecommand \@@endlink[0]{}%
\providecommand \url  [0]{\begingroup\@sanitize@url \@url }%
\providecommand \@url [1]{\endgroup\@href {#1}{\urlprefix }}%
\providecommand \urlprefix  [0]{URL }%
\providecommand \Eprint [0]{\href }%
\providecommand \doibase [0]{https://doi.org/}%
\providecommand \selectlanguage [0]{\@gobble}%
\providecommand \bibinfo  [0]{\@secondoftwo}%
\providecommand \bibfield  [0]{\@secondoftwo}%
\providecommand \translation [1]{[#1]}%
\providecommand \BibitemOpen [0]{}%
\providecommand \bibitemStop [0]{}%
\providecommand \bibitemNoStop [0]{.\EOS\space}%
\providecommand \EOS [0]{\spacefactor3000\relax}%
\providecommand \BibitemShut  [1]{\csname bibitem#1\endcsname}%
\let\auto@bib@innerbib\@empty
\bibitem [{\citenamefont {Pekola}\ \emph {et~al.}(2013)\citenamefont {Pekola}, \citenamefont {Saira}, \citenamefont {Maisi}, \citenamefont {Kemppinen}, \citenamefont {M\"ott\"onen}, \citenamefont {Pashkin},\ and\ \citenamefont {Averin}}]{Pekola_2013}%
  \BibitemOpen
  \bibfield  {author} {\bibinfo {author} {\bibfnamefont {J.~P.}\ \bibnamefont {Pekola}}, \bibinfo {author} {\bibfnamefont {O.-P.}\ \bibnamefont {Saira}}, \bibinfo {author} {\bibfnamefont {V.~F.}\ \bibnamefont {Maisi}}, \bibinfo {author} {\bibfnamefont {A.}~\bibnamefont {Kemppinen}}, \bibinfo {author} {\bibfnamefont {M.}~\bibnamefont {M\"ott\"onen}}, \bibinfo {author} {\bibfnamefont {Y.~A.}\ \bibnamefont {Pashkin}},\ and\ \bibinfo {author} {\bibfnamefont {D.~V.}\ \bibnamefont {Averin}},\ }\bibfield  {title} {\enquote {\bibinfo {title} {Single-electron current sources: Toward a refined definition of the ampere},}\ }\href {https://doi.org/10.1103/RevModPhys.85.1421} {\bibfield  {journal} {\bibinfo  {journal} {Rev. Mod. Phys.}\ }\textbf {\bibinfo {volume} {85}},\ \bibinfo {pages} {1421--1472} (\bibinfo {year} {2013})}\BibitemShut {NoStop}%
\bibitem [{\citenamefont {Kaestner}\ and\ \citenamefont {Kashcheyevs}(2015)}]{Kaestner_2015}%
  \BibitemOpen
  \bibfield  {author} {\bibinfo {author} {\bibfnamefont {B.}~\bibnamefont {Kaestner}}\ and\ \bibinfo {author} {\bibfnamefont {V.}~\bibnamefont {Kashcheyevs}},\ }\bibfield  {title} {\enquote {\bibinfo {title} {Non-adiabatic quantized charge pumping with tunable-barrier quantum dots: a review of current progress},}\ }\href {https://doi.org/10.1088/0034-4885/78/10/103901} {\bibfield  {journal} {\bibinfo  {journal} {Reports on Progress in Physics}\ }\textbf {\bibinfo {volume} {78}},\ \bibinfo {pages} {103901} (\bibinfo {year} {2015})}\BibitemShut {NoStop}%
\bibitem [{\citenamefont {Kaneko}, \citenamefont {Nakamura},\ and\ \citenamefont {Okazaki}(2016)}]{Kaneko_2016}%
  \BibitemOpen
  \bibfield  {author} {\bibinfo {author} {\bibfnamefont {N.-H.}\ \bibnamefont {Kaneko}}, \bibinfo {author} {\bibfnamefont {S.}~\bibnamefont {Nakamura}},\ and\ \bibinfo {author} {\bibfnamefont {Y.}~\bibnamefont {Okazaki}},\ }\bibfield  {title} {\enquote {\bibinfo {title} {A review of the quantum current standard},}\ }\href {https://doi.org/10.1088/0957-0233/27/3/032001} {\bibfield  {journal} {\bibinfo  {journal} {Meas. Sci. Technol.}\ }\textbf {\bibinfo {volume} {27}},\ \bibinfo {pages} {032001} (\bibinfo {year} {2016})}\BibitemShut {NoStop}%
\bibitem [{\citenamefont {Giblin}\ \emph {et~al.}(2012)\citenamefont {Giblin}, \citenamefont {Kataoka}, \citenamefont {Fletcher}, \citenamefont {See}, \citenamefont {Janssen}, \citenamefont {Griffiths}, \citenamefont {Jones}, \citenamefont {Farrer},\ and\ \citenamefont {Ritchie}}]{Giblin2012TowardsAQ}%
  \BibitemOpen
  \bibfield  {author} {\bibinfo {author} {\bibfnamefont {S.~P.}\ \bibnamefont {Giblin}}, \bibinfo {author} {\bibfnamefont {M.}~\bibnamefont {Kataoka}}, \bibinfo {author} {\bibfnamefont {J.~D.}\ \bibnamefont {Fletcher}}, \bibinfo {author} {\bibfnamefont {P.}~\bibnamefont {See}}, \bibinfo {author} {\bibfnamefont {T.~J. B.~M.}\ \bibnamefont {Janssen}}, \bibinfo {author} {\bibfnamefont {J.~P.}\ \bibnamefont {Griffiths}}, \bibinfo {author} {\bibfnamefont {G.~A.~C.}\ \bibnamefont {Jones}}, \bibinfo {author} {\bibfnamefont {I.}~\bibnamefont {Farrer}},\ and\ \bibinfo {author} {\bibfnamefont {D.~A.}\ \bibnamefont {Ritchie}},\ }\bibfield  {title} {\enquote {\bibinfo {title} {Towards a quantum representation of the ampere using single electron pumps},}\ }\href@noop {} {\bibfield  {journal} {\bibinfo  {journal} {Nat. Commun.}\ }\textbf {\bibinfo {volume} {3}} (\bibinfo {year} {2012})}\BibitemShut {NoStop}%
\bibitem [{\citenamefont {Giblin}\ \emph {et~al.}(2019)\citenamefont {Giblin}, \citenamefont {Fujiwara}, \citenamefont {Yamahata}, \citenamefont {Bae}, \citenamefont {Kim}, \citenamefont {Rossi}, \citenamefont {M{\"o}tt{\"o}nen},\ and\ \citenamefont {Kataoka}}]{Giblin2019EvidenceFU}%
  \BibitemOpen
  \bibfield  {author} {\bibinfo {author} {\bibfnamefont {S.~P.}\ \bibnamefont {Giblin}}, \bibinfo {author} {\bibfnamefont {A.}~\bibnamefont {Fujiwara}}, \bibinfo {author} {\bibfnamefont {G.}~\bibnamefont {Yamahata}}, \bibinfo {author} {\bibfnamefont {M.-H.}\ \bibnamefont {Bae}}, \bibinfo {author} {\bibfnamefont {N.}~\bibnamefont {Kim}}, \bibinfo {author} {\bibfnamefont {A.}~\bibnamefont {Rossi}}, \bibinfo {author} {\bibfnamefont {M.}~\bibnamefont {M{\"o}tt{\"o}nen}},\ and\ \bibinfo {author} {\bibfnamefont {M.}~\bibnamefont {Kataoka}},\ }\bibfield  {title} {\enquote {\bibinfo {title} {Evidence for universality of tunable-barrier electron pumps},}\ }\href@noop {} {\bibfield  {journal} {\bibinfo  {journal} {Metrologia}\ }\textbf {\bibinfo {volume} {56}} (\bibinfo {year} {2019})}\BibitemShut {NoStop}%
\bibitem [{\citenamefont {Keller}\ \emph {et~al.}(1996)\citenamefont {Keller}, \citenamefont {Martinis}, \citenamefont {Zimmerman},\ and\ \citenamefont {Steinbach}}]{Keller1996AccuracyOE}%
  \BibitemOpen
  \bibfield  {author} {\bibinfo {author} {\bibfnamefont {M.~W.}\ \bibnamefont {Keller}}, \bibinfo {author} {\bibfnamefont {J.~M.}\ \bibnamefont {Martinis}}, \bibinfo {author} {\bibfnamefont {N.~M.}\ \bibnamefont {Zimmerman}},\ and\ \bibinfo {author} {\bibfnamefont {A.}~\bibnamefont {Steinbach}},\ }\bibfield  {title} {\enquote {\bibinfo {title} {Accuracy of electron counting using a 7‐junction electron pump},}\ }\href@noop {} {\bibfield  {journal} {\bibinfo  {journal} {Appl. Phys. Lett.}\ }\textbf {\bibinfo {volume} {69}},\ \bibinfo {pages} {1804--1806} (\bibinfo {year} {1996})}\BibitemShut {NoStop}%
\bibitem [{\citenamefont {Janssen}\ and\ \citenamefont {Hartland}(2000)}]{Janssen2000AccuracyOQ}%
  \BibitemOpen
  \bibfield  {author} {\bibinfo {author} {\bibfnamefont {T.~J. B.~M.}\ \bibnamefont {Janssen}}\ and\ \bibinfo {author} {\bibfnamefont {A.}~\bibnamefont {Hartland}},\ }\bibfield  {title} {\enquote {\bibinfo {title} {Accuracy of quantized single-electron current in a one-dimensional channel},}\ }\href@noop {} {\bibfield  {journal} {\bibinfo  {journal} {Physica B}\ }\textbf {\bibinfo {volume} {284}},\ \bibinfo {pages} {1790--1791} (\bibinfo {year} {2000})}\BibitemShut {NoStop}%
\bibitem [{\citenamefont {Stein}\ \emph {et~al.}(2017)\citenamefont {Stein}, \citenamefont {Scherer}, \citenamefont {Gerster}, \citenamefont {Behr}, \citenamefont {G{\"o}tz}, \citenamefont {Pesel}, \citenamefont {Leicht}, \citenamefont {Ubbelohde}, \citenamefont {Weimann}, \citenamefont {Pierz}, \citenamefont {Schumacher},\ and\ \citenamefont {Hohls}}]{Stein2017RobustnessOS}%
  \BibitemOpen
  \bibfield  {author} {\bibinfo {author} {\bibfnamefont {F.}~\bibnamefont {Stein}}, \bibinfo {author} {\bibfnamefont {H.}~\bibnamefont {Scherer}}, \bibinfo {author} {\bibfnamefont {T.}~\bibnamefont {Gerster}}, \bibinfo {author} {\bibfnamefont {R.}~\bibnamefont {Behr}}, \bibinfo {author} {\bibfnamefont {M.}~\bibnamefont {G{\"o}tz}}, \bibinfo {author} {\bibfnamefont {E.}~\bibnamefont {Pesel}}, \bibinfo {author} {\bibfnamefont {C.}~\bibnamefont {Leicht}}, \bibinfo {author} {\bibfnamefont {N.}~\bibnamefont {Ubbelohde}}, \bibinfo {author} {\bibfnamefont {T.}~\bibnamefont {Weimann}}, \bibinfo {author} {\bibfnamefont {K.}~\bibnamefont {Pierz}}, \bibinfo {author} {\bibfnamefont {H.~W.}\ \bibnamefont {Schumacher}},\ and\ \bibinfo {author} {\bibfnamefont {F.}~\bibnamefont {Hohls}},\ }\bibfield  {title} {\enquote {\bibinfo {title} {Robustness of single-electron pumps at sub-ppm current accuracy level},}\ }\href@noop {} {\bibfield  {journal} {\bibinfo  {journal} {Metrologia}\ }\textbf {\bibinfo {volume} {54}},\ \bibinfo
  {pages} {S1 -- S8} (\bibinfo {year} {2017})}\BibitemShut {NoStop}%
\bibitem [{\citenamefont {Zhao}\ \emph {et~al.}(2017)\citenamefont {Zhao}, \citenamefont {Rossi}, \citenamefont {Giblin}, \citenamefont {Fletcher}, \citenamefont {Hudson}, \citenamefont {M\"ott\"onen}, \citenamefont {Kataoka},\ and\ \citenamefont {Dzurak}}]{Zhao2017ThermalerrorRI}%
  \BibitemOpen
  \bibfield  {author} {\bibinfo {author} {\bibfnamefont {R.}~\bibnamefont {Zhao}}, \bibinfo {author} {\bibfnamefont {A.}~\bibnamefont {Rossi}}, \bibinfo {author} {\bibfnamefont {S.~P.}\ \bibnamefont {Giblin}}, \bibinfo {author} {\bibfnamefont {J.~D.}\ \bibnamefont {Fletcher}}, \bibinfo {author} {\bibfnamefont {F.~E.}\ \bibnamefont {Hudson}}, \bibinfo {author} {\bibfnamefont {M.}~\bibnamefont {M\"ott\"onen}}, \bibinfo {author} {\bibfnamefont {M.}~\bibnamefont {Kataoka}},\ and\ \bibinfo {author} {\bibfnamefont {A.~S.}\ \bibnamefont {Dzurak}},\ }\bibfield  {title} {\enquote {\bibinfo {title} {Thermal-error regime in high-accuracy gigahertz single-electron pumping},}\ }\href {https://doi.org/10.1103/PhysRevApplied.8.044021} {\bibfield  {journal} {\bibinfo  {journal} {Phys. Rev. Appl.}\ }\textbf {\bibinfo {volume} {8}},\ \bibinfo {pages} {044021} (\bibinfo {year} {2017})}\BibitemShut {NoStop}%
\bibitem [{\citenamefont {BIPM}(2019)}]{brochure2019mise}%
  \BibitemOpen
  \bibfield  {author} {\bibinfo {author} {\bibnamefont {BIPM}},\ }\bibfield  {title} {\enquote {\bibinfo {title} {{Mise en pratique for the definition of the ampere and other electric units in the SI}},}\ }\href@noop {} {\bibfield  {journal} {\bibinfo  {journal} {SI Brochure Appendix}\ }\textbf {\bibinfo {volume} {2}},\ \bibinfo {pages} {9th} (\bibinfo {year} {2019})}\BibitemShut {NoStop}%
\bibitem [{\citenamefont {Janssen}\ \emph {et~al.}(2014)\citenamefont {Janssen}, \citenamefont {Giblin}, \citenamefont {See}, \citenamefont {Fletcher},\ and\ \citenamefont {Kataoka}}]{janssen2014redefinition}%
  \BibitemOpen
  \bibfield  {author} {\bibinfo {author} {\bibfnamefont {T.}~\bibnamefont {Janssen}}, \bibinfo {author} {\bibfnamefont {S.}~\bibnamefont {Giblin}}, \bibinfo {author} {\bibfnamefont {P.}~\bibnamefont {See}}, \bibinfo {author} {\bibfnamefont {J.}~\bibnamefont {Fletcher}},\ and\ \bibinfo {author} {\bibfnamefont {M.}~\bibnamefont {Kataoka}},\ }\bibfield  {title} {\enquote {\bibinfo {title} {Redefinition of the ampere},}\ }\href@noop {} {\bibfield  {journal} {\bibinfo  {journal} {Meas. Control.}\ }\textbf {\bibinfo {volume} {47}},\ \bibinfo {pages} {315--322} (\bibinfo {year} {2014})}\BibitemShut {NoStop}%
\bibitem [{\citenamefont {Scherer}\ and\ \citenamefont {Schumacher}(2019)}]{Scherer_2019}%
  \BibitemOpen
  \bibfield  {author} {\bibinfo {author} {\bibfnamefont {H.}~\bibnamefont {Scherer}}\ and\ \bibinfo {author} {\bibfnamefont {H.~W.}\ \bibnamefont {Schumacher}},\ }\bibfield  {title} {\enquote {\bibinfo {title} {{Single-electron pumps and quantum current metrology in the revised SI}},}\ }\href {https://doi.org/https://doi.org/10.1002/andp.201800371} {\bibfield  {journal} {\bibinfo  {journal} {Ann. Phys.}\ }\textbf {\bibinfo {volume} {531}},\ \bibinfo {pages} {1800371} (\bibinfo {year} {2019})},\ \Eprint {https://arxiv.org/abs/https://onlinelibrary.wiley.com/doi/pdf/10.1002/andp.201800371} {https://onlinelibrary.wiley.com/doi/pdf/10.1002/andp.201800371} \BibitemShut {NoStop}%
\bibitem [{\citenamefont {Maisi}\ \emph {et~al.}(2009)\citenamefont {Maisi}, \citenamefont {Pashkin}, \citenamefont {Kafanov}, \citenamefont {Tsai},\ and\ \citenamefont {Pekola}}]{maisi2009parallel}%
  \BibitemOpen
  \bibfield  {author} {\bibinfo {author} {\bibfnamefont {V.~F.}\ \bibnamefont {Maisi}}, \bibinfo {author} {\bibfnamefont {Y.~A.}\ \bibnamefont {Pashkin}}, \bibinfo {author} {\bibfnamefont {S.}~\bibnamefont {Kafanov}}, \bibinfo {author} {\bibfnamefont {J.-S.}\ \bibnamefont {Tsai}},\ and\ \bibinfo {author} {\bibfnamefont {J.~P.}\ \bibnamefont {Pekola}},\ }\bibfield  {title} {\enquote {\bibinfo {title} {Parallel pumping of electrons},}\ }\href@noop {} {\bibfield  {journal} {\bibinfo  {journal} {New J. Phys.}\ }\textbf {\bibinfo {volume} {11}},\ \bibinfo {pages} {113057} (\bibinfo {year} {2009})}\BibitemShut {NoStop}%
\bibitem [{\citenamefont {Mirovsky}\ \emph {et~al.}(2010)\citenamefont {Mirovsky}, \citenamefont {Kaestner}, \citenamefont {Leicht}, \citenamefont {Welker}, \citenamefont {Weimann}, \citenamefont {Pierz},\ and\ \citenamefont {Schumacher}}]{mirovsky2010synchronized}%
  \BibitemOpen
  \bibfield  {author} {\bibinfo {author} {\bibfnamefont {P.}~\bibnamefont {Mirovsky}}, \bibinfo {author} {\bibfnamefont {B.}~\bibnamefont {Kaestner}}, \bibinfo {author} {\bibfnamefont {C.}~\bibnamefont {Leicht}}, \bibinfo {author} {\bibfnamefont {A.}~\bibnamefont {Welker}}, \bibinfo {author} {\bibfnamefont {T.}~\bibnamefont {Weimann}}, \bibinfo {author} {\bibfnamefont {K.}~\bibnamefont {Pierz}},\ and\ \bibinfo {author} {\bibfnamefont {H.}~\bibnamefont {Schumacher}},\ }\bibfield  {title} {\enquote {\bibinfo {title} {Synchronized single electron emission from dynamical quantum dots},}\ }\href@noop {} {\bibfield  {journal} {\bibinfo  {journal} {Appl. Phys. Lett.}\ }\textbf {\bibinfo {volume} {97}} (\bibinfo {year} {2010})}\BibitemShut {NoStop}%
\bibitem [{\citenamefont {Ghee}\ \emph {et~al.}(2019)\citenamefont {Ghee}, \citenamefont {Ahn}, \citenamefont {Ryu}, \citenamefont {Sim}, \citenamefont {Hong}, \citenamefont {Kim}, \citenamefont {Bae},\ and\ \citenamefont {Kim}}]{ghee2019parallelized}%
  \BibitemOpen
  \bibfield  {author} {\bibinfo {author} {\bibfnamefont {Y.-S.}\ \bibnamefont {Ghee}}, \bibinfo {author} {\bibfnamefont {Y.-H.}\ \bibnamefont {Ahn}}, \bibinfo {author} {\bibfnamefont {S.}~\bibnamefont {Ryu}}, \bibinfo {author} {\bibfnamefont {H.-S.}\ \bibnamefont {Sim}}, \bibinfo {author} {\bibfnamefont {C.}~\bibnamefont {Hong}}, \bibinfo {author} {\bibfnamefont {B.-K.}\ \bibnamefont {Kim}}, \bibinfo {author} {\bibfnamefont {M.-H.}\ \bibnamefont {Bae}},\ and\ \bibinfo {author} {\bibfnamefont {N.}~\bibnamefont {Kim}},\ }\bibfield  {title} {\enquote {\bibinfo {title} {Parallelized single-electron pumps based on gate-tunable quantum dots},}\ }\href@noop {} {\bibfield  {journal} {\bibinfo  {journal} {J. Korean Phys. Soc.}\ }\textbf {\bibinfo {volume} {75}},\ \bibinfo {pages} {331--336} (\bibinfo {year} {2019})}\BibitemShut {NoStop}%
\bibitem [{\citenamefont {Baum}\ \emph {et~al.}(2021)\citenamefont {Baum}, \citenamefont {Amico}, \citenamefont {Howell}, \citenamefont {Hush}, \citenamefont {Liuzzi}, \citenamefont {Mundada}, \citenamefont {Merkh}, \citenamefont {Carvalho},\ and\ \citenamefont {Biercuk}}]{PRXQuantum.2.040324}%
  \BibitemOpen
  \bibfield  {author} {\bibinfo {author} {\bibfnamefont {Y.}~\bibnamefont {Baum}}, \bibinfo {author} {\bibfnamefont {M.}~\bibnamefont {Amico}}, \bibinfo {author} {\bibfnamefont {S.}~\bibnamefont {Howell}}, \bibinfo {author} {\bibfnamefont {M.}~\bibnamefont {Hush}}, \bibinfo {author} {\bibfnamefont {M.}~\bibnamefont {Liuzzi}}, \bibinfo {author} {\bibfnamefont {P.}~\bibnamefont {Mundada}}, \bibinfo {author} {\bibfnamefont {T.}~\bibnamefont {Merkh}}, \bibinfo {author} {\bibfnamefont {A.~R.}\ \bibnamefont {Carvalho}},\ and\ \bibinfo {author} {\bibfnamefont {M.~J.}\ \bibnamefont {Biercuk}},\ }\bibfield  {title} {\enquote {\bibinfo {title} {Experimental deep reinforcement learning for error-robust gate-set design on a superconducting quantum computer},}\ }\href {https://doi.org/10.1103/PRXQuantum.2.040324} {\bibfield  {journal} {\bibinfo  {journal} {PRX Quantum}\ }\textbf {\bibinfo {volume} {2}},\ \bibinfo {pages} {040324} (\bibinfo {year} {2021})}\BibitemShut {NoStop}%
\bibitem [{\citenamefont {Schuff}\ \emph {et~al.}(2024)\citenamefont {Schuff}, \citenamefont {Carballido}, \citenamefont {Kotzagiannidis}, \citenamefont {Calvo}, \citenamefont {Caselli}, \citenamefont {Rawling}, \citenamefont {Craig}, \citenamefont {van Straaten}, \citenamefont {Severin}, \citenamefont {Fedele}, \citenamefont {Svab}, \citenamefont {Kwon}, \citenamefont {Eggli}, \citenamefont {Patlatiuk}, \citenamefont {Korda}, \citenamefont {Zumbühl},\ and\ \citenamefont {Ares}}]{schuff2024fully}%
  \BibitemOpen
  \bibfield  {author} {\bibinfo {author} {\bibfnamefont {J.}~\bibnamefont {Schuff}}, \bibinfo {author} {\bibfnamefont {M.~J.}\ \bibnamefont {Carballido}}, \bibinfo {author} {\bibfnamefont {M.}~\bibnamefont {Kotzagiannidis}}, \bibinfo {author} {\bibfnamefont {J.~C.}\ \bibnamefont {Calvo}}, \bibinfo {author} {\bibfnamefont {M.}~\bibnamefont {Caselli}}, \bibinfo {author} {\bibfnamefont {J.}~\bibnamefont {Rawling}}, \bibinfo {author} {\bibfnamefont {D.~L.}\ \bibnamefont {Craig}}, \bibinfo {author} {\bibfnamefont {B.}~\bibnamefont {van Straaten}}, \bibinfo {author} {\bibfnamefont {B.}~\bibnamefont {Severin}}, \bibinfo {author} {\bibfnamefont {F.}~\bibnamefont {Fedele}}, \bibinfo {author} {\bibfnamefont {S.}~\bibnamefont {Svab}}, \bibinfo {author} {\bibfnamefont {P.~C.}\ \bibnamefont {Kwon}}, \bibinfo {author} {\bibfnamefont {R.~S.}\ \bibnamefont {Eggli}}, \bibinfo {author} {\bibfnamefont {T.}~\bibnamefont {Patlatiuk}}, \bibinfo {author} {\bibfnamefont {N.}~\bibnamefont {Korda}}, \bibinfo {author} {\bibfnamefont
  {D.}~\bibnamefont {Zumbühl}},\ and\ \bibinfo {author} {\bibfnamefont {N.}~\bibnamefont {Ares}},\ }\href@noop {} {\enquote {\bibinfo {title} {Fully autonomous tuning of a spin qubit},}\ } (\bibinfo {year} {2024}),\ \Eprint {https://arxiv.org/abs/2402.03931} {arXiv:2402.03931 [cond-mat.mes-hall]} \BibitemShut {NoStop}%
\bibitem [{\citenamefont {Lennon}\ \emph {et~al.}(2019)\citenamefont {Lennon}, \citenamefont {Moon}, \citenamefont {Camenzind}, \citenamefont {Yu}, \citenamefont {Zumb{\"u}hl}, \citenamefont {Briggs}, \citenamefont {Osborne}, \citenamefont {Laird},\ and\ \citenamefont {Ares}}]{lennon2019efficiently}%
  \BibitemOpen
  \bibfield  {author} {\bibinfo {author} {\bibfnamefont {D.~T.}\ \bibnamefont {Lennon}}, \bibinfo {author} {\bibfnamefont {H.}~\bibnamefont {Moon}}, \bibinfo {author} {\bibfnamefont {L.~C.}\ \bibnamefont {Camenzind}}, \bibinfo {author} {\bibfnamefont {L.}~\bibnamefont {Yu}}, \bibinfo {author} {\bibfnamefont {D.~M.}\ \bibnamefont {Zumb{\"u}hl}}, \bibinfo {author} {\bibfnamefont {G.~A.~D.}\ \bibnamefont {Briggs}}, \bibinfo {author} {\bibfnamefont {M.~A.}\ \bibnamefont {Osborne}}, \bibinfo {author} {\bibfnamefont {E.~A.}\ \bibnamefont {Laird}},\ and\ \bibinfo {author} {\bibfnamefont {N.}~\bibnamefont {Ares}},\ }\bibfield  {title} {\enquote {\bibinfo {title} {Efficiently measuring a quantum device using machine learning},}\ }\href@noop {} {\bibfield  {journal} {\bibinfo  {journal} {npj Quantum Information}\ }\textbf {\bibinfo {volume} {5}},\ \bibinfo {pages} {79} (\bibinfo {year} {2019})}\BibitemShut {NoStop}%
\bibitem [{\citenamefont {Giblin}\ \emph {et~al.}(2017)\citenamefont {Giblin}, \citenamefont {Bae}, \citenamefont {Kim}, \citenamefont {Ahn},\ and\ \citenamefont {Kataoka}}]{Giblin_2017}%
  \BibitemOpen
  \bibfield  {author} {\bibinfo {author} {\bibfnamefont {S.~P.}\ \bibnamefont {Giblin}}, \bibinfo {author} {\bibfnamefont {M.-H.}\ \bibnamefont {Bae}}, \bibinfo {author} {\bibfnamefont {N.}~\bibnamefont {Kim}}, \bibinfo {author} {\bibfnamefont {Y.-H.}\ \bibnamefont {Ahn}},\ and\ \bibinfo {author} {\bibfnamefont {M.}~\bibnamefont {Kataoka}},\ }\bibfield  {title} {\enquote {\bibinfo {title} {{Robust operation of a GaAs tunable barrier electron pump}},}\ }\href {https://doi.org/10.1088/1681-7575/aa634c} {\bibfield  {journal} {\bibinfo  {journal} {Metrologia}\ }\textbf {\bibinfo {volume} {54}},\ \bibinfo {pages} {299} (\bibinfo {year} {2017})}\BibitemShut {NoStop}%
\bibitem [{\citenamefont {Giblin}\ \emph {et~al.}(2020)\citenamefont {Giblin}, \citenamefont {Mykkanen}, \citenamefont {Kemppinen}, \citenamefont {Immonen}, \citenamefont {Manninen}, \citenamefont {Jenei}, \citenamefont {Mottonen}, \citenamefont {Yamahata}, \citenamefont {Fujiwara},\ and\ \citenamefont {Kataoka}}]{extrapol}%
  \BibitemOpen
  \bibfield  {author} {\bibinfo {author} {\bibfnamefont {S.~P.}\ \bibnamefont {Giblin}}, \bibinfo {author} {\bibfnamefont {E.}~\bibnamefont {Mykkanen}}, \bibinfo {author} {\bibfnamefont {A.}~\bibnamefont {Kemppinen}}, \bibinfo {author} {\bibfnamefont {P.}~\bibnamefont {Immonen}}, \bibinfo {author} {\bibfnamefont {A.}~\bibnamefont {Manninen}}, \bibinfo {author} {\bibfnamefont {M.}~\bibnamefont {Jenei}}, \bibinfo {author} {\bibfnamefont {M.}~\bibnamefont {Mottonen}}, \bibinfo {author} {\bibfnamefont {G.}~\bibnamefont {Yamahata}}, \bibinfo {author} {\bibfnamefont {A.}~\bibnamefont {Fujiwara}},\ and\ \bibinfo {author} {\bibfnamefont {M.}~\bibnamefont {Kataoka}},\ }\bibfield  {title} {\enquote {\bibinfo {title} {Realisation of a quantum current standard at liquid helium temperature with sub-ppm reproducibility},}\ }\href {https://doi.org/10.1088/1681-7575/ab72e0} {\bibfield  {journal} {\bibinfo  {journal} {Metrologia}\ }\textbf {\bibinfo {volume} {57}},\ \bibinfo {pages} {025013} (\bibinfo {year}
  {2020})}\BibitemShut {NoStop}%
\bibitem [{\citenamefont {Norimoto}\ \emph {et~al.}(2024)\citenamefont {Norimoto}, \citenamefont {See}, \citenamefont {Schoinas}, \citenamefont {Rungger}, \citenamefont {Boykin-II}, \citenamefont {Stewart-Jr}, \citenamefont {Griffiths}, \citenamefont {Chen}, \citenamefont {Ritchie},\ and\ \citenamefont {Kataoka}}]{ShotaMultiplexer}%
  \BibitemOpen
  \bibfield  {author} {\bibinfo {author} {\bibfnamefont {S.}~\bibnamefont {Norimoto}}, \bibinfo {author} {\bibfnamefont {P.}~\bibnamefont {See}}, \bibinfo {author} {\bibfnamefont {N.}~\bibnamefont {Schoinas}}, \bibinfo {author} {\bibfnamefont {I.}~\bibnamefont {Rungger}}, \bibinfo {author} {\bibfnamefont {T.~O.}\ \bibnamefont {Boykin-II}}, \bibinfo {author} {\bibfnamefont {M.~D.}\ \bibnamefont {Stewart-Jr}}, \bibinfo {author} {\bibfnamefont {J.~P.}\ \bibnamefont {Griffiths}}, \bibinfo {author} {\bibfnamefont {C.}~\bibnamefont {Chen}}, \bibinfo {author} {\bibfnamefont {D.~A.}\ \bibnamefont {Ritchie}},\ and\ \bibinfo {author} {\bibfnamefont {M.}~\bibnamefont {Kataoka}},\ }\href {https://arxiv.org/abs/2407.05926} {\enquote {\bibinfo {title} {Statistical study and parallelisation of multiplexed single-electron sources},}\ } (\bibinfo {year} {2024}),\ \Eprint {https://arxiv.org/abs/2407.05926} {arXiv:2407.05926 [cond-mat.mes-hall]} \BibitemShut {NoStop}%
\bibitem [{\citenamefont {Kaestner}\ \emph {et~al.}(2008)\citenamefont {Kaestner}, \citenamefont {Kashcheyevs}, \citenamefont {Amakawa}, \citenamefont {Blumenthal}, \citenamefont {Li}, \citenamefont {Janssen}, \citenamefont {Hein}, \citenamefont {Pierz}, \citenamefont {Weimann}, \citenamefont {Siegner},\ and\ \citenamefont {Schumacher}}]{kaestner2008}%
  \BibitemOpen
  \bibfield  {author} {\bibinfo {author} {\bibfnamefont {B.}~\bibnamefont {Kaestner}}, \bibinfo {author} {\bibfnamefont {V.}~\bibnamefont {Kashcheyevs}}, \bibinfo {author} {\bibfnamefont {S.}~\bibnamefont {Amakawa}}, \bibinfo {author} {\bibfnamefont {M.~D.}\ \bibnamefont {Blumenthal}}, \bibinfo {author} {\bibfnamefont {L.}~\bibnamefont {Li}}, \bibinfo {author} {\bibfnamefont {T.~J. B.~M.}\ \bibnamefont {Janssen}}, \bibinfo {author} {\bibfnamefont {G.}~\bibnamefont {Hein}}, \bibinfo {author} {\bibfnamefont {K.}~\bibnamefont {Pierz}}, \bibinfo {author} {\bibfnamefont {T.}~\bibnamefont {Weimann}}, \bibinfo {author} {\bibfnamefont {U.}~\bibnamefont {Siegner}},\ and\ \bibinfo {author} {\bibfnamefont {H.~W.}\ \bibnamefont {Schumacher}},\ }\bibfield  {title} {\enquote {\bibinfo {title} {Single-parameter nonadiabatic quantized charge pumping},}\ }\href {https://doi.org/10.1103/PhysRevB.77.153301} {\bibfield  {journal} {\bibinfo  {journal} {Phys. Rev. B}\ }\textbf {\bibinfo {volume} {77}},\ \bibinfo {pages} {153301}
  (\bibinfo {year} {2008})}\BibitemShut {NoStop}%
\bibitem [{\citenamefont {Howe}\ \emph {et~al.}(2021)\citenamefont {Howe}, \citenamefont {Blumenthal}, \citenamefont {Beere}, \citenamefont {Mitchell}, \citenamefont {Ritchie},\ and\ \citenamefont {Pepper}}]{howe2021single}%
  \BibitemOpen
  \bibfield  {author} {\bibinfo {author} {\bibfnamefont {H.}~\bibnamefont {Howe}}, \bibinfo {author} {\bibfnamefont {M.}~\bibnamefont {Blumenthal}}, \bibinfo {author} {\bibfnamefont {H.}~\bibnamefont {Beere}}, \bibinfo {author} {\bibfnamefont {T.}~\bibnamefont {Mitchell}}, \bibinfo {author} {\bibfnamefont {D.}~\bibnamefont {Ritchie}},\ and\ \bibinfo {author} {\bibfnamefont {M.}~\bibnamefont {Pepper}},\ }\bibfield  {title} {\enquote {\bibinfo {title} {Single-electron pump with highly controllable plateaus},}\ }\href@noop {} {\bibfield  {journal} {\bibinfo  {journal} {Appl. Phys. Lett.}\ }\textbf {\bibinfo {volume} {119}} (\bibinfo {year} {2021})}\BibitemShut {NoStop}%
\bibitem [{\citenamefont {Timbers}, \citenamefont {Campbell},\ and\ \citenamefont {Lee}(2022)}]{timbers2022data}%
  \BibitemOpen
  \bibfield  {author} {\bibinfo {author} {\bibfnamefont {T.}~\bibnamefont {Timbers}}, \bibinfo {author} {\bibfnamefont {T.}~\bibnamefont {Campbell}},\ and\ \bibinfo {author} {\bibfnamefont {M.}~\bibnamefont {Lee}},\ }\href@noop {} {\emph {\bibinfo {title} {Data science: A first introduction}}}\ (\bibinfo  {publisher} {Chapman and Hall/CRC},\ \bibinfo {year} {2022})\BibitemShut {NoStop}%
\bibitem [{\citenamefont {Aly}, \citenamefont {Aref},\ and\ \citenamefont {Ouzzani}(2015)}]{aly2015cost}%
  \BibitemOpen
  \bibfield  {author} {\bibinfo {author} {\bibfnamefont {A.~M.}\ \bibnamefont {Aly}}, \bibinfo {author} {\bibfnamefont {W.~G.}\ \bibnamefont {Aref}},\ and\ \bibinfo {author} {\bibfnamefont {M.}~\bibnamefont {Ouzzani}},\ }\bibfield  {title} {\enquote {\bibinfo {title} {Cost estimation of spatial k-nearest-neighbor operators.}}\ }in\ \href@noop {} {\emph {\bibinfo {booktitle} {EDBT}}}\ (\bibinfo {year} {2015})\ pp.\ \bibinfo {pages} {457--468}\BibitemShut {NoStop}%
\bibitem [{\citenamefont {Pedregosa}\ \emph {et~al.}(2011)\citenamefont {Pedregosa}, \citenamefont {Varoquaux}, \citenamefont {Gramfort}, \citenamefont {Michel}, \citenamefont {Thirion}, \citenamefont {Grisel}, \citenamefont {Blondel}, \citenamefont {Prettenhofer}, \citenamefont {Weiss}, \citenamefont {Dubourg}, \citenamefont {Vanderplas}, \citenamefont {Passos}, \citenamefont {Cournapeau}, \citenamefont {Brucher}, \citenamefont {Perrot},\ and\ \citenamefont {Duchesnay}}]{scikit-learn}%
  \BibitemOpen
  \bibfield  {author} {\bibinfo {author} {\bibfnamefont {F.}~\bibnamefont {Pedregosa}}, \bibinfo {author} {\bibfnamefont {G.}~\bibnamefont {Varoquaux}}, \bibinfo {author} {\bibfnamefont {A.}~\bibnamefont {Gramfort}}, \bibinfo {author} {\bibfnamefont {V.}~\bibnamefont {Michel}}, \bibinfo {author} {\bibfnamefont {B.}~\bibnamefont {Thirion}}, \bibinfo {author} {\bibfnamefont {O.}~\bibnamefont {Grisel}}, \bibinfo {author} {\bibfnamefont {M.}~\bibnamefont {Blondel}}, \bibinfo {author} {\bibfnamefont {P.}~\bibnamefont {Prettenhofer}}, \bibinfo {author} {\bibfnamefont {R.}~\bibnamefont {Weiss}}, \bibinfo {author} {\bibfnamefont {V.}~\bibnamefont {Dubourg}}, \bibinfo {author} {\bibfnamefont {J.}~\bibnamefont {Vanderplas}}, \bibinfo {author} {\bibfnamefont {A.}~\bibnamefont {Passos}}, \bibinfo {author} {\bibfnamefont {D.}~\bibnamefont {Cournapeau}}, \bibinfo {author} {\bibfnamefont {M.}~\bibnamefont {Brucher}}, \bibinfo {author} {\bibfnamefont {M.}~\bibnamefont {Perrot}},\ and\ \bibinfo {author} {\bibfnamefont
  {E.}~\bibnamefont {Duchesnay}},\ }\bibfield  {title} {\enquote {\bibinfo {title} {Scikit-learn: Machine learning in {P}ython},}\ }\href@noop {} {\bibfield  {journal} {\bibinfo  {journal} {J. Mach. Learn. Res.}\ }\textbf {\bibinfo {volume} {12}},\ \bibinfo {pages} {2825--2830} (\bibinfo {year} {2011})}\BibitemShut {NoStop}%
\bibitem [{\citenamefont {Fukuda}(2004)}]{fukuda2004frequently}%
  \BibitemOpen
  \bibfield  {author} {\bibinfo {author} {\bibfnamefont {K.}~\bibnamefont {Fukuda}},\ }\bibfield  {title} {\enquote {\bibinfo {title} {Frequently asked questions in polyhedral computation},}\ }\href {https://www.cs.mcgill.ca/~fukuda/soft/polyfaq/polyfaq.html} {\  (\bibinfo {year} {2004})}\BibitemShut {NoStop}%
\bibitem [{\citenamefont {Barber}, \citenamefont {Dobkin},\ and\ \citenamefont {Huhdanpaa}(2013)}]{barber2013qhull}%
  \BibitemOpen
  \bibfield  {author} {\bibinfo {author} {\bibfnamefont {C.~B.}\ \bibnamefont {Barber}}, \bibinfo {author} {\bibfnamefont {D.~P.}\ \bibnamefont {Dobkin}},\ and\ \bibinfo {author} {\bibfnamefont {H.}~\bibnamefont {Huhdanpaa}},\ }\href@noop {} {\enquote {\bibinfo {title} {Qhull: Quickhull algorithm for computing the convex hull},}\ } (\bibinfo {year} {2013})\BibitemShut {NoStop}%
\bibitem [{\citenamefont {Virtanen}\ \emph {et~al.}(2020)\citenamefont {Virtanen}, \citenamefont {Gommers}, \citenamefont {Oliphant}, \citenamefont {Haberland}, \citenamefont {Reddy}, \citenamefont {Cournapeau}, \citenamefont {Burovski}, \citenamefont {Peterson}, \citenamefont {Weckesser}, \citenamefont {Bright} \emph {et~al.}}]{virtanen2020scipy}%
  \BibitemOpen
  \bibfield  {author} {\bibinfo {author} {\bibfnamefont {P.}~\bibnamefont {Virtanen}}, \bibinfo {author} {\bibfnamefont {R.}~\bibnamefont {Gommers}}, \bibinfo {author} {\bibfnamefont {T.~E.}\ \bibnamefont {Oliphant}}, \bibinfo {author} {\bibfnamefont {M.}~\bibnamefont {Haberland}}, \bibinfo {author} {\bibfnamefont {T.}~\bibnamefont {Reddy}}, \bibinfo {author} {\bibfnamefont {D.}~\bibnamefont {Cournapeau}}, \bibinfo {author} {\bibfnamefont {E.}~\bibnamefont {Burovski}}, \bibinfo {author} {\bibfnamefont {P.}~\bibnamefont {Peterson}}, \bibinfo {author} {\bibfnamefont {W.}~\bibnamefont {Weckesser}}, \bibinfo {author} {\bibfnamefont {J.}~\bibnamefont {Bright}}, \emph {et~al.},\ }\bibfield  {title} {\enquote {\bibinfo {title} {Scipy 1.0: fundamental algorithms for scientific computing in python},}\ }\href@noop {} {\bibfield  {journal} {\bibinfo  {journal} {Nat. Methods}\ }\textbf {\bibinfo {volume} {17}},\ \bibinfo {pages} {261--272} (\bibinfo {year} {2020})}\BibitemShut {NoStop}%
\bibitem [{\citenamefont {Chen}(2017)}]{chen2017tutorial}%
  \BibitemOpen
  \bibfield  {author} {\bibinfo {author} {\bibfnamefont {Y.-C.}\ \bibnamefont {Chen}},\ }\bibfield  {title} {\enquote {\bibinfo {title} {A tutorial on kernel density estimation and recent advances},}\ }\href@noop {} {\bibfield  {journal} {\bibinfo  {journal} {Biostat. \& Epidemiol.}\ }\textbf {\bibinfo {volume} {1}},\ \bibinfo {pages} {161--187} (\bibinfo {year} {2017})}\BibitemShut {NoStop}%
\bibitem [{\citenamefont {W\k{e}glarczyk}(2018)}]{wkeglarczyk2018kernel}%
  \BibitemOpen
  \bibfield  {author} {\bibinfo {author} {\bibfnamefont {S.}~\bibnamefont {W\k{e}glarczyk}},\ }\bibfield  {title} {\enquote {\bibinfo {title} {Kernel density estimation and its application},}\ }in\ \href@noop {} {\emph {\bibinfo {booktitle} {ITM web of conferences}}},\ Vol.~\bibinfo {volume} {23}\ (\bibinfo {organization} {EDP Sciences},\ \bibinfo {year} {2018})\ p.\ \bibinfo {pages} {00037}\BibitemShut {NoStop}%
\end{thebibliography}%

\end{document}